\documentclass[twocolumn,aps,prl,10pt,amsmath,amssymb,nofootinbib,showpacs,superscriptaddress,floatfix]{revtex4-1}

\DeclareFontFamily{U}{rcjhbltx}{}
\DeclareFontShape{U}{rcjhbltx}{m}{n}{<->rcjhbltx}{}
\DeclareSymbolFont{hebrewletters}{U}{rcjhbltx}{m}{n}

\usepackage{graphicx}
\usepackage{color}
\usepackage[usenames,dvipsnames]{xcolor}
\usepackage[colorlinks=true,linkcolor=Red,citecolor=Green,linktoc=page]{hyperref}
\usepackage{multirow}
\usepackage{float}
\usepackage{flushend}
\usepackage{balance}
\usepackage[varg]{txfonts}
\usepackage{ulem}
\usepackage{fancyhdr}

\DeclareMathSymbol{\lamed}{\mathord}{hebrewletters}{108}

\begin{document}
\title{Mirror modular cloning and fast quantum associative retrieval }

\author{M.\, C.\, Diamantini}

\affiliation{NiPS Laboratory, INFN and Dipartimento di Fisica e Geologia, University of Perugia, via A. Pascoli, I-06100 Perugia, Italy}

\author{C.\,A.\,Trugenberger}

\affiliation{SwissScientific Technologies SA, rue du Rhone 59, CH-1204 Geneva, Switzerland.}

\begin{abstract}
We show that a quantum state can be perfectly cloned up to global mirroring with a unitary transformation that depends on one single parameter. We then show that this is equivalent to ``perfect" cloning for quantum associative memories which, as a consequence efficiently hold exponentially more information than their classical counterparts. Finally, we present a quantum associative retrieval algorithm which can correct corrupted inputs and is exponentially faster than the Grover algorithm. 

\end{abstract}
\maketitle


The no-cloning theorem \cite{zurek} is a cornerstone of quantum information theory. On one side it implies that information cannot be intercepted without detection, which is the basis of the widely extolled power of quantum cryptography, on the other side it hampers the reproduction of information stored in quantum states for multiple use, i.e. the use of quantum memories. 

While, initially quantum computation was hailed mainly for the speed-up with respect to its classical counterpart (for a review see \cite{nielsen}), it was soon realized that quantum mechanics also leads to an exponential increase in storage capacity for associative retrieval of information \cite{cat1, cat2, ventura} and an exponential decrease in memory calls for quantum random access memories (RAM) \cite{lloyd1}. The prototype associative memory is the Hopfield neural network model \cite{hopfield}. Due to the cross-talk phenomenon, the maximum number of memories that can be classically stored and retrieved is linear in the number $n$ of neurons (for a review see \cite{muller}). The quantization of the Hopfield model mapping neurons to qubits does not improve on this limit \cite{dt}. The exponential gain in capacity is obtained when the patterns are stored in quantum states \cite{cat1, ventura, lloyd2}. 

The original idea \cite{cat1, ventura} is to store a number $p$ of binary patterns of $n$ qubits in a quantum superposition and use this state as the memory. Two different algorithms for retrieval were proposed, one associative, i.e. content-addressable \cite{cat1}, the other based on the Grover algorithm \cite{grover} using a subset of the $n$ qubits as the memory address \cite{ventura}. The latter is essentially a quantum RAM memory since it cannot correct corrupted inputs. Various variants of the quantum associative memories have subsequently been proposed \cite{petruccione, sousa, khan}. Recently, the whole field of machine learning is being extended to the quantum domain \cite{briegel, biamonte, havlicek}. 

In an ideal situation one would simply clone the quantum memory state whenever needed for use and keep a master copy for later re-use. However, the no-cloning theorem prevents this. In this paper we point out that, for quantum memories, perfect cloning of a single state is not necessary, it is sufficient to clone the memory state up to a global NOT operation that transforms it into its mirror image. This {\it mirror modular cloning} can be performed by a (2 $\times $ 2) unitary transformation that depends on a single parameter of the state and represents thus an easy and efficient ``perfect" cloning of the memory state. We then introduce an improved version of the quantum associative recall and show that this is exponentially faster than the address-based Grover retrieval. This is particularly important in view of recent proposals to use quantum memories for fast associative data triggering in large throughput high-energy experiments at LHC \cite{lhc,lhc2}. 

Let us start from a quantum memory state $|M \rangle $ encoding $p$ binary patterns $|p^i \rangle = |p^i_1 \dots p^i_n\rangle $ of $n$ qubits,
\begin{equation}
|M\rangle = {1\over \sqrt{p}} \sum_{i=1}^p |p^i \rangle \ .
\label{memory}
\end{equation}
An efficient algorithm to load the patterns to form this state is described in \cite{cat1}, here we will not repeat the procedure but rather take this state as our starting point. The same algorithm can be used to construct the mirror modular state
\begin{equation}
|\overline M\rangle = {1\over \sqrt{p}} \sum_{i=1}^p | \overline p^i \rangle \ ,
\label{mirrorstate}
\end{equation}
where $\overline p^i $ is the string in which each qubit is reversed compared to $p^i$. Finally, we measure the scalar product 
$\langle M | \overline M \rangle$ (see for e.g. \cite{stolze}) of these two states. 

We now add a normalized ancillary register of $n$ qubits prepared in state $|\Sigma \rangle$ and a further ancilla qubit in state $|0\rangle$ and we borrow a technique from probabilistic quantum cloning \cite{guo} to posit a $2 \times 2$ unitary transformation such that
\begin{eqnarray}
U\left( |M\rangle |\Sigma \rangle |0\rangle \right) &&= \sqrt{\gamma}  |M\rangle |M\rangle |0\rangle + \sqrt{\overline \gamma} |M\rangle |\overline M\rangle |1\rangle  \ ,
\nonumber \\
U\left( |\overline M\rangle |\Sigma \rangle |0\rangle \right) &&= \sqrt{\overline \gamma}  |\overline M\rangle |\overline M\rangle |0\rangle + \sqrt{ \gamma} |\overline M\rangle |M\rangle |1\rangle  \ .
\label{posited}
\end{eqnarray} 
This unitary transformation, if it exists, perfectly clones the memory state up to a mirror modular transformation, outputting $|M\rangle$ with probability $\gamma$ or $|\overline M\rangle$ with probability $\overline \gamma$, the two results being distinguished by the value of the ancilla qubit. As we will show below, this mirror modular cloning is perfectly sufficient for quantum associative retrieval. Before doing that, however, we must show that $U$ exists. 

To do so we use the theorem stating that there exists a unitary transformation $U|\psi_i \rangle = |\phi_i\rangle$, $i = 1 \dots m$, if the 
two sets of states $|\psi_i\rangle$ and $|\phi_i\rangle$ satisfy $\langle \psi_i | \psi_j \rangle = \langle \phi_i | \phi_j \rangle$ for all $i, j = 1\dots m$. The $2\times 2$ matrix of all scalar products of initial states in (\ref{posited}) is
\begin{equation}
\begin{pmatrix}
1 & \langle M | \overline M \rangle \\
\langle \overline M | M\rangle & 1	 
\end{pmatrix}
\label{inmatrix}
\end{equation}
The corresponding matrix of scalar products of unitary transformed states is
\begin{equation}
\begin{pmatrix}
\gamma + \overline \gamma & \sqrt{\gamma \overline \gamma} (\langle M | \overline M \rangle  +  \langle \overline M| M \rangle) \langle M| \overline M\rangle \\
 \sqrt{\gamma \overline \gamma} (\langle \overline M | M \rangle  + \langle M| \overline M \rangle) \langle \overline M|M\rangle & \gamma + \overline \gamma 	 
\end{pmatrix}
\label{outmatrix}
\end{equation}
This shows that the unitary transformation $U$ exists if the efficiencies $\gamma $ and $\overline \gamma$ are chosen to satifsy
\begin{eqnarray}
\gamma + \overline \gamma &&= 1 \ ,
\nonumber \\
\sqrt{\gamma \overline \gamma} &&= {1\over \langle M | \overline M \rangle  +  \langle \overline M| M \rangle} \ .
\label{eff}
\end{eqnarray}
It can then be realized with a quantum circuit along the lines detailed in \cite{circuit}. 

After separating out the original master copy of the memory state we add an $n$ qubit register with an input pattern $|I\rangle = |i_1 \dots i_n \rangle$ and we add $b$ control qubits prepared in state $ |O\rangle = |c_1 \dots c_b \rangle = |0,\dots 0\rangle$. This is the initial state for the associative retrieval algorithm,
\begin{equation}
|\psi_0 \rangle = \sqrt{\gamma} |I\rangle |M\rangle |O\rangle |0\rangle + \sqrt{\overline \gamma}|I\rangle |\overline M\rangle |O\rangle |1\rangle \ .
\label{initial}
\end{equation} 
Before describing retrieval in detail let us focus on the elementary quantum gates (for a review see \cite{nielsen}) used in this algorithm. First of all there are the single-qubit NOT gate, represented by the first Pauli matrix $\sigma_1$, and Hadamard gate H, with the matrix representation
\begin{equation}
H = {1\over \sqrt{2}} 
\begin{pmatrix} 
1 & 1\\
1 & -1 
\end{pmatrix} \ .
\label{adda}
\end{equation}
Moreover, we will use the two-qbit XOR (exclusive OR) gate, which performs a NOT operation on the second qubit if and only if the first one is in state $|1\rangle$. In matrix notation this gate is represented as ${\rm XOR} = {\rm diag} \left( I, \sigma_1 \right)$, where $I$ denotes a two-dimensional identity matrix. For all these gates we shall indicate by subscripts the qubits on which they are applied, the control qubits coming always first.  

We start the retrieval algorithm by generating the state
\begin{eqnarray}
|\psi _1\rangle &&= \prod_{k=1}^n \ {\rm NOT}_{m_k} 
\ {\rm XOR}_{i_k m_k} |\psi _0\rangle 
\nonumber \\
&&= \sqrt{\gamma \over p} \sum_{k=1}^p |I\rangle |D^k\rangle |O\rangle |0\rangle + \sqrt{\overline \gamma\over p} \sum_{k=1}^p |I\rangle |\overline D^k\rangle|O\rangle  |1\rangle \ ,
\label{second}
\end{eqnarray}
where $|D^k\rangle = |d^k_1 \dots d^k_n \rangle $ and $|\overline D^k\rangle = |\overline d^k_1 \dots \overline d^k_n \rangle $ with $d^k_j=1$ iff $p^k_j = i_j$ and $\overline d^k_j=1$ iff $p^k_j \ne i_j$. We now consider the unitary transformation
\begin{equation} 
U =  \prod_{i=1}^b H_{c_i} \  {\rm e}^{ i\pi {\cal H}_i/2n} \ H_{c_i} \ ,
\label{uham}
\end{equation}
where ``$c_i$" refers to the ${\rm i}^{\rm th}$ control qubit and
\begin{eqnarray}
{\cal H}_i &&= \left( d_H \right)_m \otimes \left( \sigma_3 \right)_{c_i} \ ,
\nonumber \\
\left( d_H \right)_m && = \sum_{k=1}^n 
\left( {\sigma_3 + 1\over 2} \right) _{m_k}\ ,
\label{h}
\end{eqnarray}
with $\sigma _3$ the third Pauli matrix. For each control qubit ${\cal H}_i$ measures the number of 0's in the memory register, now in state $D^K$ or $\overline D^k$, with a plus sign if $c_i$ is in state $|0\rangle$ and a minus sign if $c_i$ is in state $|1\rangle$. When the memory register is in state $|D^k\rangle$ (first term in (\ref{second})) this is the number of qubits which are different in the input and memory registers. This quantity is called the {\it Hamming distance} $d_H\left (i, p^k \right)$ between the stored pattern $p^k$ under consideration and the input $i$. When the memory register is in state $|\overline D^k\rangle$ (second term in (\ref{second})), instead, it represents the number of qubits which are equal in the input and the memory registers, which we denote by $\overline d_H\left (i, p^k \right)$ when the memory register contains pattern $p^k$. 

The two Hadamard gates sandwiching the Hamming distance operator in (\ref{uham}) turn every control qubit into the combination
\begin{equation}
{\rm cos}  {\pi \over 2n} d_H \left( i, p^k \right) |0\rangle + i {\rm sin} {\pi \over 2n} d_H \left( i, p^k \right) |1\rangle \ ,
\label{cossin}
\end{equation}
in the term corresponding to distance pattern $|D^k\rangle$, with an analogous expression in terms of $\overline d_H$ for distance patterns $|\overline D^k\rangle$. After restoring the memory register to its original state with the inverse transformation to (\ref{second}), we obtain the state 
\begin{eqnarray}
|\psi_2\rangle &&= \sqrt{\gamma \over p} \sum_{k=1}^p \sum_{l=0}^b
\ {\rm cos}^{b-l} \left( {\pi\over 2n} d_H\left( i, p^k \right)\right) \times 
\nonumber \\
&&({\rm i \ sin})^l \left( {\pi\over 2n} d_H\left( i, p^k \right)\right) 
\ \sum_{\left\{ J^l \right\}} |I\rangle |p^k \rangle |J^l\rangle |0\rangle 
\nonumber \\
&&+\sqrt{\overline \gamma \over p} \sum_{k=1}^p \sum_{l=0}^b
\ {\rm cos}^{b-l} \left( {\pi\over 2n} \overline d_H\left( i, p^k \right)\right) \times 
\nonumber \\
&&({\rm i \ sin})^l \left( {\pi\over 2n} \overline d_H\left( i, p^k \right)\right) 
\ \sum_{\left\{ J^l \right\}} |I\rangle |\overline p^k \rangle |J^l\rangle |1\rangle  \ ,
\label{al}
\end{eqnarray}
where $\left\{ J^l \right\}$ denotes the set of all binary numbers of $b$ bits with exactly $l$ bits 1 and $(b-l)$ bits 0. At this point we measure the original ancilla qubit: if $|0\rangle $ is obtained we use amplitude amplification \cite{brassard} to rotate the remaining state onto the ``good" subspace with all control qubits in state $|0_1 \dots 0_b\rangle$, if $|1\rangle$ is obtained we use the same technique to rotate the remaining state onto the ``good" subspace with all control qubits in state $|1_1 \dots 1_b\rangle$. After these rotations we can finally measure the memory register. We either obtain the closest (in Hamming distance) pattern $p^k$ (if $|0\rangle$ was measured) or the most distant pattern $\overline p^k$ (if $|1\rangle$ was measured) with a probability distribution
\begin{equation}
P \left (i, p^k \right) = {1\over p} {\rm cos}^{2b} \left( {\pi \over 2n} d_H \left(i, p^k \right) \right) \ .
\label{prob}
\end{equation} 
The probability distribution is the same since, for mirror patterns, $\overline d_H = n-d_H$ and ${\sin} \left( (\pi/2) (1-d_H/n) \right) = {\rm cos} \left( \pi d_H/2n \right)$. This shows that, as anticipated, it does not matter if one obtains the closest or the most distant pattern: if the original ancilla qubit is measured in state $|1\rangle$ one needs only to mirror-invert the measured pattern. This is why 
memory collapse and the difficulty of cloning do not pose any problem for quantum associative memories, differently to what is sometimes discussed in the literature \cite{petruccione, briegel, sousa, khan}. Mirror modular cloning is a simple and efficient technique that depends on only one parameter of the state. 

Finally, the complexity of the retrieval algorithm is mainly influenced by the amplitude amplification step. The number of applications $C$ of the basic amplitude amplification rotation is given by the square root of the inverse probability of measuring the ``good" subspace, i.e. 
\begin{equation}
C = \sqrt{{p \over \sum_{k=1}^p {\rm cos}^{2b} \left( {\pi \over 2n} d_H \left(i, p^k \right) \right)}} \ .
\label{compl1}
\end{equation}
This is not a known quantity since it depends on the input under consideration. So, either we need C repetitions of the algorithm, we use amplitude estimation \cite{brassard} or we record an estimate of C together with the initial memory storage by computing it for some typical inputs, after which we can repeat amplitude amplification a few times varying the number of oracle calls around this initial estimate. Compared to the complexity $\sqrt{N}$ to retrieve one pattern with the address-oriented Grover algorithm \cite{ventura}, where $N$ is the total number of computational basis states, $N=2^n$, the present complexity not only does not depend on $N$, but it does not even depend on the pattern number $p$ since it is determined only by the accuracy parameter $b$ and the distribution of patterns in the memory. For an approximately uniform pattern distribution it can be estimated as follows. Let us substitute each cosine in the sum by its average, determined using
\begin{equation}
{2\over \pi} \int_0^{\pi/2} {\rm cos}^{2b} (x) \ dx = {1\over 2^{2b}} {2b \choose b} \ ,
\label{average}
\end{equation}
and let us approximate the factorials by the Stirling formula $b! \approx \sqrt{2\pi b} (b/e)^b$ for large $b$. With this we obtain
\begin{equation}
C \approx (\pi b)^{1/4} \ ,
\label{compl2}
\end{equation}
which shows that the present quantum associative retrieval offers an exponential speed-up with respect to the Grover-based one, contrary to what asserted in \cite{lhc}. Of course there is a trade-off between complexity and accuracy: the higher $b$ the more the retrieval probability (\ref{prob}) is peaked on the correct pattern but the higher the computational complexity. 

In conclusion, the quantum associative memory originally proposed in \cite{cat1}, with the improvements proposed here, is not limited by the no-cloning theorem, can correct corrupted inputs on top of completing partial ones and offers an exponential speed-up with respect to the Grover-based alternative.


\begin{thebibliography}{10}
	\expandafter\ifx\csname url\endcsname\relax
	\def\url#1{\texttt{#1}}\fi
	\expandafter\ifx\csname urlprefix\endcsname\relax\def\urlprefix{URL }\fi
	\providecommand{\bibinfo}[2]{#2}
	\providecommand{\eprint}[2][]{\url{#2}}
	
\bibitem{zurek}W. Wootters and W. Zurek, A single quantum cannot be cloned, {\it Nature} {\bf 299}, 802-803 (1982).

\bibitem{nielsen} M. A. Nielsen and I. L. Chuang, Quantum computation and quantum information, Cambridge University Press, Cambridge (200). 

\bibitem{cat1} C. A. Trugenberger, Probabilistic quantum memories, {\it Phys. Rev. Lett} {\bf 89} 067901 (2001) 

\bibitem{cat2} C. A. Trugenberger, Quantum pattern recognition, {\it Quantum Inf. Process.} {\bf 1} 471 (2002). 

\bibitem{ventura} D. Ventura and T. Martinez, Quantum associative memory {\it Found. Phys. Lett.} {\bf 12} 547 (1999). 

\bibitem{lloyd1}V. Giovannetti, S. Lloyd and L. Maccone, Quantum random access memory, {\it Phys. Rev. Lett.} {\bf 100} 160501 (2008).

\bibitem{hopfield} J. J. Hopfield, Neural networks and physical systems with emergent collective computational abilities, {\it Proceeding of the national academy of science USA} {\bf 79} 2554 (1982).

\bibitem{muller} B. M\"uller and J. Reinhardt, Neural networks, Springer-Verlag, Berlin (1990). 

\bibitem{dt} M. C. Diamantini and C. A. Trugenberger, Quantum pattern retrieval by qubit networks with Hebb interactions, {\it Phys. Rev. Lett.} {\bf 97} 130503 (2006). 

\bibitem{lloyd2} P. Rebentrost, T. R. Bromley, C. Weedbrook and S. Lloyd, Quantum Hopfield neural network, {\it Phys. Rev. } {\bf A98} 042308 (2018). 

\bibitem{grover}L. Grover, Quantum mechanics helps in searching for a needle in a haystack, {\it Phys. Rev. Lett.} {\bf 79} 325 (1997). 

\bibitem{petruccione} M. Schuld, I. Sinayskiy and F. Petruccione, Quantum computing for pattern classification, {\it Pacific Rim International Conference on Artificial Intelligence} Springer-Verlag, 208-220 (2014). 

\bibitem{sousa} R. S. Sousa, P. G. M. dos Santos, T. M. L. Veras, W. R. de Oliveira and A. J. da Silva, Parametric probabilistic quantum memory, {\it Neurocomputing} {\bf 416} 360-369 (2020).

\bibitem{khan} M. Khan, J. P. L. Faye, U. C. Mendes and A. Miransky, EP-PQM: efficient parametric probabilistic quantum memory with fewer qubits and gates, {\it IEEE Transactions on Quantum Engineering} (2022). 

\bibitem{briegel}V. Dunjko and H. J. Briegel, Machine learning and artificial intelligence in the quantum domain: a review of recent progress, {\it Reports on Progress in Physics} {\bf 81} 074001 (2018).

\bibitem{biamonte} J. Biamonte, P. Wittek, N. Pancotti. P. Rebentrost and N. Wiebe, Quantum machine learning, {\it Nature} {\bf 549} 195-202 (2017). 

\bibitem{havlicek} V. Havlicek, A. D. Corcoles, K. Temme, A. W. Harrow and A. Kandala, Supervised learning with quantum-enhanced feature spaces, {\it Nature} {\bf 567} 209-212 (2019). 

\bibitem{lhc} I. Shapoval and P. Calafiura, Quantum associative memory in HEP track pattern recognition, {\it EPJ Web of Conferences} {\bf 214} 01012 (2019). 

\bibitem{lhc2}H. M. Gray, Quantum pattern recognition algorithms for charged particle tracking, {\it Phil. Trans. R. Soc. A} {\bf 380} 20210103 (2021). 

\bibitem{stolze}J. Stolze and A. I. Zenchuk, Computing scalar products via a two-terminal quantum transmission line, {\it Phys. Lett. A} {\bf 383} 125978 (2019). 

\bibitem{guo} L.-M. Duan and G.-C. Guo, Probabilistic cloning and identification of linearly independent quantum states, {\it Phys. Rev. Lett.} {\bf 80} 4999 (1998). 

\bibitem{circuit}C.-W. Zhang, Z.-Y. Wang, C.-F. Li, G.-C. Guo, Realizing probabilistic identification and cloning of quantum states via universal quantum logic gates, {\it Phys. Rev. } {\bf A61} 062310 (2000). 

\bibitem{brassard} G. Brassard, P. Hoyer, M. Mosca and A. Tapp, Amplitude amplification and estimation, {\it Contemporary Mathematics} {\bf 305} 53-74 (2002). 


	
\end{thebibliography}
\end{document}